\documentclass[runningheads,a4paper]{llncs}

\usepackage{amssymb}
\usepackage{graphicx}
\usepackage{url}
\usepackage{amsfonts}

\font\dsr=dsrom10

\def\Z{\mbox{\dsr Z}}

\usepackage{bbm}
\usepackage[greek,english]{babel}
\usepackage{ucs}
\usepackage[utf8x]{inputenc}
\usepackage{autofe}

\DeclareUnicodeCharacter{8759}{\ensuremath{::}}

\usepackage{fancyvrb}

\DefineVerbatimEnvironment
   {code}{Verbatim}
   {} 

\begin{document}

\title{On dependent types and intuitionism\\ in programming mathematics}

\titlerunning{Dependent types and intuitionism in programming}

\author{Sergei D. Meshveliani
        \thanks{This work is supported by the FANO project of the Russian Academy of
                Sciences, the project registration No AAAA-A16-116021760039-0.}
       } 
\institute{Program Systems Institute of Russian Academy of sciences,\\ 
           Pereslavl-Zalessky, Russia. 
           email: mechvel@botik.ru
          }
\maketitle

\begin{abstract}
It is discussed a practical possibility of a provable programming of mathematics
basing on intuitionism and the dependent types feature of a programming language.
The principles of constructive mathematics and provable programming are illustrated
with examples taken from algebra. The discourse follows the experience in designing 
in {\tt Agda} a computer algebra library {\tt DoCon-A}, which deals with
generic algebraic structures and also provides the needed machine-checked proofs. 

This paper is a revised translation of a certain paper published in Russian in 2014.  
\end{abstract}



\section{Introduction}
\label{sec-intro}

This paper is a certain revised translation of the paper \cite{Me1} published in 
Russian in 2014.  

It describes the experience in searching for the most appropriate tool for programming
mathematical computation. The investigation is based on the practice of programming 
symbolic computations in algebra.

The goal was to find an universal programming language with possibly rich 
mathematical expressiveness, to explain its advantages with respect to other 
languages, and to test this tool on implementing a considerable piece of a real-world 
computer algebra.

The history of this search and program design experience consists of the three main 
steps. They correspond to the three language classes:
\begin{enumerate}
\item generic programming languages,
\item purely functional languages,
\item languages with dependent types.
\end{enumerate}

Below it is explained of why the feature (1) is important.
For example, the {\tt Haskell} language \cite{Ha} belongs to the classes (1) and (2).
In this language the author has implemented in 1990-es the {\tt DoCon} library 
for commutative algebra \cite{Me2} \cite{Me3}.  
It is presumed here that the notion of pure functionality, and the main features of 
the {\tt Haskell} language, are known to the reader.

Finally, we consider the {\tt Agda} language \cite{Ag} \cite{Nor:Cha}, which belongs 
to the classes (1), (2), (3). Developing a workable implementation for such a complex
language is a great technical problem. Currently, {\tt Agda} is, mainly, a workable 
tool (with a great field remaining for desirable optimizations).
In this paper we try to show that {\tt Agda} fits best the needs of programming 
computation  in mathematics.

\emph{About the sources on mathematics:} the mathematical notions used in this paper can 
be found in \cite{Wa} (algebra) and in \cite{Dav}, \cite{CA-p} (computer algebra).

The following discourse concerns mainly the ways to program mathematical computations and 
proofs in the {\tt Agda} language.

\subsection{Generic programming} 

Programs for mathematical computation often operate in different domains in a common way.
For example, algebra textbooks present a simple algorithm to compute the greatest common 
divisor (gcd) for a pair of elements in any Euclidean ring E. There is an infinite set of
domains that can be substituted for E. Such are: integer number domain ℤ, 
the domain {\tt ℚ[x]} of univariate polynomials with rational coefficients,   
the domain {\tt (ℤ/(p))[x]} of univariate polynomials with coefficients modulo a prime 
integer {\tt p}, and infinitely many other domains.   

This approach (generic programming) is implemented in programming languages: 
there have appeared languages with abstract/polymorphic types, with type classes. 
In such a language a program like gcd is written once for all Euclidean rings.
For example, in the {\tt Haskell} language each class of algebraic domains 
(we call it here a \emph{generic algebraic structure})
(\emph{group, ring, field}, and such) can be expressed as a \emph{type class}, 
and each concrete domain of this class is called an \emph{instance} of this class.
A generic structure (roughly speaking) pretends to be an abstract theory, 
and its instance pretends to be a model of this theory.

For example, a generic structure of groups by addition can be (partially) expressed 
by the declaration

{\footnotesize
\begin{code}
  class Group a where (+) :: a -> a -> a
                      0   :: a
                      neg :: a -> a
\end{code}
}
\medskip

\noindent
Note that this is only a signature of a generic structure, other group laws cannot be 
expressed in {\tt Haskell}. 

Further, if the types {\tt a} and {\tt b} have instances of {\tt Group}, 
then it is possible to define a {\tt Group} instance for the direct product 
{\tt (a, b)} by the laws of direct product of groups:

{\footnotesize
\begin{code}
  instance (Group a, Group b) => Group (a, b)
                                 where
                                 (x, y) + (x', y') = (x + x', y + y')
                                 0                 = (0, 0)
                                 neg (x, y)        = (neg x, neg y)
\end{code}
}
\medskip

\noindent
This pretends to be the functor of direct group product.

In some other languages (say, {\tt ML}) there are possible other constructs to 
support generic programming.

The first profound approach with generic programming in mathematics was implemented
in 1977 -- 1990 in the {\tt Spad} language and in the {\tt Axiom} library for scientific
computation \cite{Axi} written in {\tt Spad}. 
Further, it has been designed the {\tt Aldor} language, as a refinement and extension
for {\tt Spad}, it even has the dependent types feature.

The {\tt DoCon} library \cite{Me2} \cite{Me3} is written in {\tt Haskell}.
The most fundamental difference points of this language from {\tt Aldor} are    
pure functionality, the ``lazy'' computation model, absence of dependent types
(the last feature is negative).

There was an additional reason for designing the {\tt DoCon} library in {\tt Haskell}: 
the commercial status of the {\tt Axiom} system in 1990-es.

\subsection{The problem of a domain depending on a value}
\label{sec-dynamParam}

Consider the example: the domain \ {\tt D = ℤ/(m)} \ of integers modulo {\tt m}
depends on the parameter {\tt m}. There are known various computation methods
for {\tt D}, in which {\tt m} is changed at run-time over the set of values which 
is not even known at the stage of compilation. Also the correctness condition 
of a method may depend on {\tt m}. For example, the Gauss method to solve a linear
system over {\tt D} is correct only for a prime {\tt m}. 

Another example: the domain of integer matrices {\tt m × n} is a \emph{semigroup} 
by the matrix multiplication only if {\tt m = n}. Generally, the set of valid 
instances for a domain often depends on a value being computed at run-time.

The type systems of the languages like {\tt Haskell} and {\tt ML} do not provide 
constructs to represent such domains in a fully adequate way.

On the other hand, in a mathematical textbook, it is possible to describe a 
computation that operates, for example, with the domain \ {\tt D = ℤ/(m)}, \ 
with changing the value of {\tt m} in a loop, and applying different methods 
depending on whether {\tt m} is prime or not.
So that we see a certain inadequacy of the type systems. 

To handle with this inadequacy, the {\tt DoCon} library applies an explicit coding 
of a domain as an {\tt Haskell} data --- in addition to the set of instances
related to the type. This evil complication is the cost paid for mathematical 
expressiveness of the library.

The approach of interpreter of the domain coding in checking the correctness 
conditions for a domain leads to that 1) this check is programmed by the library 
implementor, or by the user (not by the compiler), which approach is not reliable 
2) this leads to that the condition break for a method may become detected only 
after many hours of computation.

The approach with dependent types differs in that the type checker operates
with the type expressions like {\tt ℤ/(m)} above by doing certain symbolic 
computation on such expressions at compile-time, with treating {\tt m} as a variable.
For example, the primality condition for {\tt m} can be is expressed as a certain
type depending on {\tt m}.  
So that this way of processing type expressions corresponds adequately to 
setting algorithms in the mathematical textbooks as described above.

\subsection{Dependent types. The investigation goal}
\label{sec-goal2}

There have been designed programming languages with dependent types, in which the 
above adequacy problem is solved: {\tt Aldor} \cite{Ald}, {\tt Gallina} --- the 
language of the {\tt Coq} system \cite{Coq}, {\tt Agda} \cite{Ag} \cite{Nor:Cha}.
In these languages a type may depend of a value, and computation with types 
can be programmed as well as with ordinary values (at least this is so in {\tt Agda}).

In 2013, the author started to design the version {\tt DoCon-A} of the {\tt DoCon}
library that bases on the dependent types feature, with using the {\tt Agda} 
language. {\tt Agda} is chosen due to the following reasons.

\begin{itemize}
\item It is purely functional.
\item It has the ``lazy'' computation model. 
\item It is close to {\tt Haskell} (roughly speaking, it is an extension of 
      {\tt Haskell}), and the previous library is written in {\tt Haskell}, 
  as well as a certain prover, which waits for its application to composing 
  proofs in {\tt Agda} programs.
\item Such is a personal taste and experience of the library developer.  
\end{itemize}

Initially, the library author considered dependent types only due to the problem
of a domain depending on a value. But then, it occurs that this also brings in a 
powerful tool of a logical language.

Dependent types give the three main advantages:
\begin{itemize}
\item adequate representation of a domain depending on values,
\item automatic verification,
\item the possibility to formulate notions that are ``understood'' by the type checker,
      and to include in a program statements about this notions, and formal proofs for
      this statements, which proofs are checked automatically.   
\end{itemize}

The goal of the project is to 
\begin{itemize}
\item rewrite the {\tt DoCon} library in the {\tt Agda} language,
\item investigate on this example the practical possibilities of the approach 
      of constructive mathematics and provable programming based on dependent types.
\end{itemize}

Of course, the part of ``to rewrite'' meets various problems related to the 
approach of constructive mathematics (intuitionism) \cite{Mar} \cite{Loe}.  

In the sequel the approach of intuitionism and dependent types is explained on examples,
and there are described some design principles for the library 
(\cite{Me4} and the manual of \cite{Me6} give much more detail). 

Below the {\tt DoCon-A 2.00} library is often called {\tt DoCon-A}, or `library'.

\section{Some {\tt Agda} features. Simple program examples}
\label{sec-someFeaturesExamples}

We need to explain certain lexical and graphical features of the language, because
without this most {\tt Agda} programs are not possible to percept.

The {\tt Agda} parser reads symbols in the UTF-8 coding. This makes it possible to
set to the program such identifiers as, for example, 
ℕ, ≈, ≉, ≡, ≰, ∘, ∙, x⁻¹, ¬.

The UTF symbols are entered to the source via the text editor under the editor 
mode {\tt agda-mode}. In this mode the editor shows correctly these symbols on the 
screen.

Names (identifiers) in a program are separated by blank, for example operator or a 
variable is separated by a blank.

\emph{Example:} consider the code fragment

{\footnotesize
\begin{code}
  m : ℕ
  m = foo1
  n = foo2

  2*n≥m : 2 * n ≥ m   -- declaration of membership to a type
  2*n≥m = f m n       -- applying function  f                                  

  p = g 2*n≥m         -- applying g to a value in the type (2 * n ≥ m)         
\end{code}
}
\medskip

\noindent 
Here \ \ {\tt 2*n≥m} \ \ is a variable. It denotes a value in the type \
{\footnotesize {\tt 2 * n ≥ m}}. \
And the symbols \ {\tt `*', `≥'} \ in the expression of this type denote respectively
an operator and a data constructor --- because they are separated with blanks.

Note also that the name \ {\tt 2*n≥m} \ of a variable is self-explaining, the reader
can guess from it what is the type of this value, and what is its meaning.

So: the lexical rules of {\tt Agda} bring in \ a) special mathematical symbols, \
b) more mnemonic and sense for identifiers.

\subsection{Example of a program: forming a rational number} 
\label{sec-someFeatures:formingRational}

Let a function \ {\footnotesize {\tt f : ℕ → ℕ}} \ implement a map of natural numbers.
And consider the function
{\footnotesize
\begin{code}
  g : ℕ → Rational
  g n =  record{ num = 1;  denom = f n;  denom≢0 = f<n>≢0 }
         where
         f<n>≢0 =  <proof>
\end{code}
}
\medskip

\noindent 
that uses {\tt f} for forming a rational number \ {\footnotesize {\tt 1/(f n)}}.

The language construct {\tt `:'} means ``(this value) belongs to this type''.

{\tt `T → U'} \ means the type of all functions from the type {\tt T} to the type {\tt U}.

The type for rational numbers can be declared as the following \emph{record}:
{\footnotesize
\begin{code}
  record Rational : Set where
                        field num      : ℕ
                              denom    : ℕ
                              denom/=0 : ¬ (denom ≡ 0)
\end{code}
}
\noindent
In this record, the fields {\tt num} and {\tt denom} are respectively the parts of
numerator and denominator, {\tt denom≢0} is the field for a witness (proof) for that 
{\tt denom} is not equal zero (otherwise the fraction is not defined).

The term {\tt (denom ≡ 0)} is a certain \emph{type family} {\tt T},
depending of a parameter (``index'') {\tt denom}. Any data \ {\tt d : T} of this 
type is a witness for the statement \ {\tt ``denom = 0''}.

All this means that the above line \ {\tt g n = \ldots} \ implements forming of a
rational number \linebreak {\footnotesize {\tt 1/(f n)}}.

In the function {\tt g}, the part \verb#<proof># needs to be a function (algorithm)
that forms any element {\tt ne} of type \ {\footnotesize {\tt nT = ¬ (f n ≡ 0)}} \ 
(that is a proof for negation of the equality of \ {\footnotesize {\tt (f n) ≡ 0})}. 
This formal proof uses the definition of the relation {\tt ≡} (in Standard library), 
definition for the function {\tt f} (which we omit), and it expresses a proof for the 
statement \ {\footnotesize {f n ≢ 0}}. It can be built (in the user program), for example, by 
induction by {\tt n}.

The type checker checks that the algorithm for the {\tt ne} value returns a data 
in the type {\tt nT}. This is done by a certain symbolic computation with type
terms (normalization of a term by the program equations, unification of terms, 
and such).  

This check is done before run-time, it does not need giving {\tt n} a concrete 
value. And this all expresses forming a proof for a certain statement in which
{\tt n} is under the universal quantifier.

\subsection{Constructive version for logical connectives}
\label{sec-someFeatures:connectives}

In constructive logic, implication is represented as a map (algorithm) from 
some type {\tt T} to some type {\tt U} --- a map between two domains of witnesses.
If the types {\tt T} and {\tt U} represent the statements {\tt P} and {\tt Q}
respectively, then the type \ {\tt T → U} \ represents the implication \ 
\verb#P => Q#. \ Because any function {\tt f} of type {\tt T → U} maps any 
witness {\tt t : T} for {\tt P} to a witness \ {\tt (f t) : U} \ for {\tt Q}.

The constructive conjunction of {\tt P} and {\tt Q} is represented as a direct 
product {\tt T × U} of the types. A member of this product is represented as a
pair data {\tt (t , u)}.

The constructive disjunction of {\tt P} and {\tt Q} is represented as the disjoint 
union of the types: \linebreak {\tt T $⊎$ U}. \  
a member of this union is written as the data {\tt (inj₁ t)} or {\tt (inj₂ u)}, 
where {\tt inj₁} and {\tt inj₂} are the constructors of imbedding into the union 
from the types {\tt T} and {\tt U} respectively.

× and $⊎$ are not language constructs, they are type constructors defined in Standard 
library (implemented in {\tt Agda}).

The following example gives a more definite notion of provable programming with
using dependent types.

\subsection{Example: defining a semigroup}
\label{sec-someFeatures:semig}

The semigroup notion can be expressed as 
{\footnotesize
\begin{code}
  record Semigroup (A : Setoid) : Set
    where
    open Setoid A using (_≈_) renaming (Carrier to C)
    field
      _∙_   : C → C → C
      cong∙ : (x y x' y' : C) → x ≈ x' → y ≈ y' → (x ∙ y) ≈ (x' ∙ y')
      assoc : (x y z : C) → (x ∙ y) ∙ z ≈ x ∙ (y ∙ z)
\end{code}
}
\medskip

\noindent 
Here {\tt Setoid} is any set {\tt Carrier} with any equivalence relation \verb#_≈_# 
defined on it. This definition is taken from Standard library. The relation
\verb#_≈_# is implemented by the programmer for each particular setoid instance.

A semigroup is a setoid on which carrier {\tt C} it is defined a binary operation
\verb#_∙_# satisfying the laws of \verb#cong∙, assoc#.

The field {\tt cong∙} means the congruence law --- of that the relation {\tt \_≈\_} 
and the operation {\tt \_∙\_} agree.
This property is expressed as a type of a function. Each function of this type 
(this needs to be an algorithm accompanied by a termination proof) is a proof
for this law for the corresponding semigroup.
The expression of this type contains the indices {\footnotesize {\tt x y x' y'}}.  
The function {\tt cong∙} maps each such quadruple and witnesses for \  
{\footnotesize {\tt x ≈ x', y ≈ y'}} \ into a proof for the equality \linebreak 
{\footnotesize {\tt (x ∙ y) ≈ (x' ∙ y')}}.

The signature for {\tt cong∙} is a constructive representation for the statement 
\begin{center}
  for all {\footnotesize {\tt x, y, x', y'}} from {\tt C} 
          (if {\footnotesize {\tt x ≈ x'}} and {\footnotesize {\tt y ≈ y'}}, \ then \ 
                                               {\footnotesize {\tt x ∙ y ≈ x' ∙ y'}}).
\end{center}

\noindent
The types constituting the signature for the {\tt cong∙} value depend on the 
values {\footnotesize {\tt x, y, x', y'}}.
This allows to represent the statement about the operation {\tt \_∙\_} and relation
{\tt \_≈\_} in the form of a certain type family.

Also it is used here a constructive representation for implication via the type 
constructor {\tt \_→\_} --- as it is described in previous section.

The {\tt assoc} field has the type of a witness for associativity for {\tt \_∙\_}.

\subsection{Example: a semigroup of natural numbers}
\label{sec-someFeatures:semigNat}

It is given a program defining a notion of semigroup.
And the semigroup instance for natural numbers is implemented by the following 
program:

{\footnotesize 
\begin{code}
  nat+semigroup : Semigroup Nat.setoid
  nat+semigroup =
                  record{ _∙_ = _+_;  cong∙ = cong+;  assoc = assoc+ }
      where
      _+_ : ℕ → ℕ → ℕ                    -- addition in unary system
      0       + n = n
      (suc m) + n = suc (m + n)

      _=n_ =  Setoid._≈_ Nat.setoid      -- equality on ℕ          

      assoc+ : (x y z : ℕ) → (x + y) + z =n x + (y + z)
      assoc+ 0       y z =  refl
      assoc+ (suc x) y z =
                       begin ((suc x) + y) + z    =n[ refl ]
                             suc ((x + y) + z)    =n[ PE.cong suc (assoc+ x y z) ]
                             suc (x + (y + z))    =n[ refl ]
                             (suc x) + (y + z)
                       □

      cong∙ =  < skip >
\end{code}
}
\medskip

\noindent Let us comment this.

Natural numbers (of type ℕ) are written in unary coding, via the data constructor
{\tt suc} (``successor'').

{\tt Nat.setoid} is the setoid instance for ℕ imported from Standard library. 
                 Its carrier is ℕ.

The semigroup operation \verb#_∙_# is implemented as the addition \verb#_+_# on ℕ.
The two equations for \verb#_+_# implement the algorithm to evaluate this operation.

But the type checker needs to check termination of the algorithm. It this case,
it is done by a certain built-in procedure. In the second equation the first 
argument {\tt m} for the operation {\tt +} in the right-hand side is syntactically 
smaller then the first argument {\footnotesize {\tt (suc m)}} for {\tt +} in the 
left-hand side. Hence, the type checker concludes that this recursion terminates. 

The function {\tt assoc+} is a proof for associativity for the operation {\tt +}. 
It is done by induction by the construction of the first argument. 
In the first equation, the data constructor {\tt refl} means that the equality \ 
{\footnotesize {\tt (0 + y) + z =n 0 + (y + z)}} \ is proved by reducing its parts to 
the normal form according to the definition of the function {\tt +}. This is a computation
on the type terms, with presence of variables ({\tt y, z}). If the two normal 
forms do not coincide, the type checker reports of that the {\tt refl} data is not 
a needed proof. Otherwise the final proof is by applying the reflexivity law \linebreak 
{\footnotesize {\tt X =n X}}.

In the second equation for {\tt assoc+}, the right-hand side is a proof for the 
equality \linebreak {\footnotesize \verb#((suc x) + y) + z  =n  (suc x) + (y + z)#}. \ 
It is represented by the three successive equality transformations.
In each line, the construct {\tt =n[ \ldots ]}  presents a composition of the 
laws which prove the equality of the value in the current line to the value in the 
next line. 

Thus, the construct \ {\footnotesize {\tt =n[ PE.cong suc (assoc+ x y z) ]}} \ 
denotes that the associativity law is applied to the subterm \ 
{\footnotesize {\tt (x + y) + z}} \ 
in the current term, and then it is applied the congruence law for the {\tt suc}
constructor with respect to the equality {\tt =n}.

\subsubsection{Digression about operation congruence.}

Note that in algebra we always deal with a \emph{theory with equality}.
For example, if {\footnotesize {\tt x ≈ y}}, then \ 
{\footnotesize {\tt f x ≈ f y, \ x + a ≈ y + a}},  \ldots

In most textbooks on mathematics this congruence law is presumed. 
But an {\tt Agda} program needs to point explicitly, where congruence holds,
and to operate explicitly with its witnesses (as it is shown above in the 
construct of \ {\footnotesize {\tt PE.cong suc (\ldots)}}). Otherwise the type checker would
not accept a proof. And this is natural, because in an arbitrary program the 
result is not necessarily agreed with the instance for the equality {\tt ≈}
used in this program, taking also in account that {\tt ≈} is most often implemented
by the programmer.

Let us return to the last proof in the example.
In the construct of

\begin{center}
              {\footnotesize \verb#(begin ...   =n[ ...  ]  ... □)# }
\end{center}
\verb#begin_#  is a certain function applied prefixly,
\verb#_□# a function applied postfixly,
\verb#_=n[_]_#  in an infix function of three arguments
(it is a function from Standard library renamed to \verb#_=n[_]_#, we skip this import
in the above code).

The construct of \ {\footnotesize \verb#(begin ...   =n[ ...  ]  ... □)# } \ 
is not a language construct. This is only applying the three functions programmed in 
{\tt Agda} in Standard library.

This approach with introducing for proofs certain functions with infix denotation 
has the effect of implementing various languages for proofs.

{\tt Agda} has not any special language for proofs. Proofs are written in the same 
language as all the rest.

A considerable experience of the {\tt DoCon-A} library \cite{Me6} in composing 
proofs (see the manual) does not show the necessity of any special language for 
proofs. Still there is needed further experience to decide on this.

\section{The {\tt DoCon-A} library as implementation
         of a part of constructive mathematics}
\label{sec-partOfConsMath}

Constructive mathematics \cite{Mar} puts that each object must be built by 
some given algorithm accompanied by a proof for termination. For example, the 
{\tt DoCon} library implements the algorithms for linear system solving, finding
Gr\"obner basis (\cite{CA-p}, Appendix I), factoring a polynomial over certain
coefficient domains, and many others. 
For all these methods, as well as for many other useful methods, composing a 
formal termination proof does not present any real problem.

\subsection{About searching for a constructive proof}
\label{sec-partOfConsMath:consProof}

\subsubsection{Termination by syntactic decrement.} 

In numerous easy cases, the type checker derives termination itself by 
observing a recursive call and finding an argument that becomes syntactically 
smaller, in a certain appropriate ordering (similar as with the above rules for 
{\tt +}).

\subsubsection{Termination by counter.} 

In many other cases this built-in proof does not work. The library uses in such 
cases introducing to the function an additional argument --- 
a counter for ``steps'' of type ℕ (with the {\tt suc} constructor). 
And a termination proof is obtained via comparison of the counter value with 
an appropriate size function value for a certain argument, when the counter 
decreases by one occurrence of {\tt suc} with each recursion step.
  
(Proofs of this kind are often given in textbooks and papers).

Also there is a feature of \emph{sized types}, but we do not consider here
this tool.

\subsubsection{Termination by unfeasible bound.} 

Sometimes it is known that the given algorithm terminates, but a proof for 
termination is rather complex, this may be a solution for some great problem.
And it is needed to apply this algorithm without setting its termination proof
to the program (which proof may take, say, a thousand of source pages).

There is an elegant way out. Add to the function a counter argument {\tt cnt} for 
the number of steps, and a bound {\tt B} for {\tt cnt}, so that {\tt cnt} is 
decreased from {\tt B} towards {\tt 0}. And put {\tt B} to be some
\emph{unfeasible} number, for example, \verb#2 ^ (10 ^ 100)#.  
Then a termination proof for the modified algorithm is simple, and the two 
algorithms produce the same result for all input for which the number of steps 
is feasible. 

(The trick with unfeasible bound is known to me from an e-mail message by Ulf Norell).

\subsubsection{Termination for semidecision.} 

Various semidecision algorithms are sometimes useful in practice. For example,
searching a proof for an equation for complex enough calculus.
A way out for this may be adding a bound for the number of steps.
Both bounds may have sense, a feasible bound and unfeasible one.

\subsubsection{Postulating termination.} 

If the programmer is lazy to design some messy termination proof, one can 
delay this proof for future by postulating termination for a particular function.
This is done by setting the  \verb& {-# TERMINATING #-} &  pragma strictly before 
the function declaration.

{\tt DoCon-A} uses the {\tt TERMINATING} pragma only in certain two places
where such usage is not important. These places are normal form functions in 
{\tt EqProver/*} and in some {\tt Read} instances. The excuse for this is as follows.
\begin{itemize} 
\item A proof for this termination can be provided in future.
\item The {\tt EqProver} functions run only at the stage of type checking, they are 
      parts of so-called \emph{tactics}.
\item The {\tt Read} instances is a new feature, currently it has a draft implementation.
\end{itemize}

\subsubsection{Non-constructive existence.} 

Some computations and proofs in mathematics may include non-constructive operations.
Consider the discourse
``As $R$ is a Noetherian ring with unity, there exists a maximal ideal $I$ in $R$, and
this ideal is generated by some finite generator set $G$.  
Then, put the result $h = f(G)$''. In classical mathematics, the above maximal ideal 
$I$ does exist. But choosing such an ideal cannot be done by an algorithm, in general 
case. 

As an example of most easy constructivity problem among the difficult ones, 
consider the Higman's lemma \cite{Hig}:

\begin{itemize}
\item[] For any infinite sequence $w(k)$ of words in a finite alphabet there exist 
        numbers $i$ and $j$ such that $w(i)$ is a subword in $w(j)$\\
        (that is $w(i)$ is obtained from $w(j)$ by deleting several positions,
         may be, no positions).
\end{itemize}

\noindent
It is known (C.~Nash-Williams) its short and non-constructive proof. And it was 
expected that a constructive proof will be much more complex and long. Later there
have been obtained machine-checked proofs with using the systems
{\tt Coq, Isabelle}, and finally {\tt Agda}: \cite{Ber} \cite{Rom}.
Thus, the proof in {\tt Agda} (for the case of a two letters alphabet) takes
only about 80 non-empty lines of the source program.

Finally: the last tool to use for handling any problems with constructivity is the 
`{\tt postulate}' construct.

{\tt DoCon-A} does not use such.

\subsection{Proof by contradiction}

\emph{Very often it is possible in constructive mathematics}
(and in the {\tt Agda} language). This is so, for example, in such a case when 
the corresponding relation {\tt P} is decidable --- there is given an algorithm
to solve {\tt P}.

The library contains many proofs by contradiction for decidable relations. 
An important source of this decidability is decidability of the equality
relation ≈, which is required for classical generic structures. 

As an example of undecidable relation, consider the equality relation in some groups
defined by several generators and quotiented by several equalities (the word problem
in a finitely generated group).

In principle, it is possible to postulate the law of \emph{excluded third} 
and to apply it in {\tt Agda} proofs, relying on the classical logic. 

But for a system dealing with algorithms, it is much more appropriate to keep 
constructivity as far as possible.

\section{Summary: about advantages of provable dependently-typed programming}

The theoretical base for programming with dependent types is the intuitionistic 
type theory by M. Loef \cite{Loe}.

As soon as dependent types are used, algorithms (and program) are joined with 
proofs. And this makes it possible a provable programming, when the principal 
properties of the algorithm are automatically checked by the type checker.

More definitely, dependent types provide the following possibilities. 

\begin{enumerate}
\item To express a property {\tt P} of the algorithm as a type {\tt T}
      (depending on values), where the data constructors for {\tt T} can be defined
      by the programmer.

\item To express a proof for {\tt P} as a function that builds any value in {\tt T}.

\item To join in the source program the algorithm and a proof for its main properties
      (chosen by the programmer), and to do it so that this does not effect the 
      run-time performance,

\item To rely on automatic check of the proofs.

\item To automatically check many theorem proofs in mathematics (for statements
      are expressed as dependent types).
\end{enumerate}

Also the last two points are important because 
\begin{itemize}
\item most good textbooks have errors (which are often caused by typos),
\item an error in a program that drives a device may cause heavy effects.
\end{itemize}

Let us sketch certain important features of formal proofs in {\tt Agda}.

\begin{enumerate}
\item The type checker would not accept an erroneous on incomplete proof.

\item A proof is a \emph{data} of the {\tt Agda} language, and sometimes it  
      has sense for a program to analyze the structure of the proof.

\item The type checker first searches for a proof by default, by normalizing
  the type expressions by the definitions of functions in the scope.
  For example, the type\\ 
  {\footnotesize {\tt (suc 0) + (suc (suc 0)) ≡ (suc (suc 0)) + (suc 0)}} \ \ 
  is normalized by the Standard library functions to \ \ 
  {\footnotesize {\tt suc (suc (suc 0)) ≡ suc (suc (suc 0))}}, \ \ 
  and the latter has a proof given by the standard constructor {\tt refl}.

  Proof by normalization often occurs sufficient, and even more often occurs not
  sufficient.

\item The {\tt `postulate'} construct can be set for a proof that the programmer
  is lazy to compose. The means ``trust me, so far''.
  Even if all properties are postulated, this still produces a program in which 
  domains are represented more adequately than in a language without dependent types.

\item There are some functions from Standard library which help composing proofs,
      Also most Standard library functions are provided with the corresponding proofs.

\item It is desirable to add to the library more provers (written in {\tt Agda})
      which help composing proofs.

\item A proof in {\tt Agda} program is formal and complete. A proof of a statement 
    is represented as an implementation for the function having the goal signature.
    This function is formed as a composition of functions which are proofs for some 
    lemmata.
\end{enumerate}

\subsection{Example: a program for sorting a list}

This function has an additional argument: a decidable ordering structure for the 
domain of the list elements.
An usual approach in dependently-typed programming is the following. 

\begin{enumerate}
\item \label{(1)} The notions of the relation \verb#_<_# being a total order
  and \verb#_<?_# being the corresponding decision for \verb#_<_# are written in 
  the form of a type declaration. All this is written as a record of type 
  {\tt DecTotalOrder}.

\item The notion of a list being ordered by the relation \verb#_<_# is written
      in the form of a type declaration.

\item It is written an algorithm {\tt sort} for sorting. 
      The sorting function is applied as \verb#(sort dto xs#),  
  where the record \ {\tt dto : DecTotalOrder} \ contains the instances of \ 
  \verb#<, <?# \ and witnesses for their above properties.

\item The function sort returns the record of type \ {\tt SortResult xs} \ 
  which has the fields of {\tt ys, ord-ys, mSetEq}. \ 
  {\tt ys} is the result list, {\tt ord-ys} is a witness for that {\tt ys} 
  is ordered, {\tt mSetEq} is a witness for that {\tt xs} and {\tt ys} have the 
  same multiset of elements.
\end{enumerate}

See the file {\tt List/Sorting.agda} in {\tt DoCon-A} where it is programmed the 
merge method for sorting.

Let us note that without dependent types the property of the relation \verb#_<_# 
being a total order cannot be expressed, neither it is checked by the compiler. 
And in the case of unlucky implementation for \verb#_<_#, for example, this relation 
may occur not transitive, and the result list may occur not ordered.

\section{The current state of the {\tt DoCon-A} project}
\label{sec-currentState}

The current {\tt DoCon-A 2.00} release \cite{Me6} implements only a small subset
of the \emph{methods} from the {\tt DoCon} library, this is a certain introduction
to the future provable version of {\tt DoCon}. But this introduction includes an 
adequate domain representation and full machine-checked justification of all the 
used algorithms and constructs. And this makes it a large program which tests on practice
the possibility to express a real-world computer algebra library in {\tt Agda}.

{\tt DoCon-A 2.00} \cite{Me6} implements the following hierarchy of classical algebraic 
structures: 

\newpage
{\footnotesize                                                         
\begin{code}                                                                       
        DSet       (a set with decidable equality _≈_, with conditional enumeration)
        |                                              
    *<- Magma      (a set with abstract binary operation _∙_)            
    |   |                                                        
-<--|-- Semigroup  --> CommutativeSemigroup           
|   |   |                                                 
|   |   Monoid     --> CommutativeMonoid        
|   |   |              |                   
|   |   |              CCMonoid              (with the cancellation law)
|   |   |              |                                                
|   |   Group          FactorizationMonoid   (with factoring to primes) 
|   |   |                                
|   |   CommutativeGroup                          
|   |   |                                   
|    -> Ringoid                
|       |                
*-----> Ring                    
        |                        
        RingWithOne      
        |                  
        CommutativeRing            
        |                                              
        IntegralRing                         (∀ x y → (x*y ≈ 0 → x≈0 or y≈0))  
        |       \   \      
        |        \    -----  EuclideanRing   (division with remainder, etc.)  
        |         \                                  
        |           -------  GCDRing         (with an algorithm for gcd)       
        |                                                     
        FactorizationRing                    (with an algorithm to factor to primes)
        |                 
        Unique FactorizationRing             (the prime∣split property added)   
        |                          
        Field                                (each nonzero has an inverse)


Figure 1.  The tower of classical algebraic structures supported in DoCon-A-2.00.
\end{code}                                     
}
\bigskip

The following features are implemented.

\begin{itemize}
\setlength\itemsep{0pt}
\item The \emph{domain constructors} of \ {\tt Natural (ℕ), Integer (ℤ)}, \ 
      direct product for semigroups, \ 
      {\tt Fraction, \ UnivariatePolynomial, \ EuclideanResidue}.

\item For ℕ there are implemented the instances of \
      {\tt CommutativeMonoid} for addition and multiplication, and
      {\tt UniqueFactorizationMonoid} for multiplication for \ ℕ\verb#\0# \cite{Me5}.

\item For ℤ there are implemented the instances of
      {\tt EuclideanRing} and {\tt UniqueFactorizationRing}.

\item For {\tt Fraction} (over any unique factorization ring) \ there is
      implemented the {\tt Field} instance, with certified optimized methods for
      arithmetic.

\item An univariate {\tt Polynomial} over any {\tt CommutativeRing} \
      is represented by a certain pair list ordered decreasingly by exponents.
  The corresponding \verb#_+_# operation is implemented, and there are proved
  its properties of associativity and commutativity.

\item {\tt EucResidue} is the constructor of the residue ring {\tt R/(b)} for any
      Euclidean ring {\tt R} and any its nonzero element {\tt b} being not
  invertible. This constructor builds the instance of {\tt CommutativeRing} in the
  general case, and it builds the instance of {\tt Field} when {\tt b} is detected
  as prime.

\item The demonstration program {\tt demoTest/EucResTest.agda} runs the examples
      of arithmetics in {\tt R/(b)} for the instance of {\tt R =} $\Z$.

\item The extended GCD method is programmed for an arbitrary {\tt EuclideanRing}.

\item A rich sub-library ({\tt AList}) is implemented for operations with lists,
    association, lists, multisets. It includes the merge sorting function with all the
    needed proofs.

\item The notions of {\tt FactorizationMonoid, FactorizationRing,
      FactorizationIsUnique}  are defined, and there are proved many lemmata about 
      them \cite{Me5}.

\item For the binary-coded natural numbers ({\tt Bin}) it is added a proof for
      bijectiveness of the coding ({\tt toDigits}) and for its homomorphism with
      respect to the successor operation.

\item It is implemented the binary method for powering in a monoid, with proofs.
      It uses the binary coding for the exponent.

\item Certain special equational provers \ {\tt InMonoid, InSemiringWithOne},\\
      {\tt InCommutativeSemiring} \ are implemented and used.

\item All the classical definitions --- properties are formulated as types for
    the above notions and constructs (as in a textbook on algebra, only given in full),
    all the proofs are provided for the methods.

\item The performance of the programs {\tt demoTest/EucResTest} \
      (for residue ring),\\ {\tt demoTest/SortingTest} (for merge sorting),
      {\tt demoTest/FractionTest} (for fraction arithmetic) is nearly as in the 
      {\tt DoCon} system (under Glasgow Haskell).
\end{itemize}
\medskip

Let us consider some details.

\subsection{Computational cost of a proof}
\label{sec-currentState:proofCost}

Proofs do not effect the run-time performance of a program, for a reasonably designed
program. But they take 
\begin{itemize}
\item a volume of the program source code,
\item memory space and processor time at the stage of type checking,
\item time and effort in composing proofs, currently it is great.
\end{itemize}

\noindent
For example, type-checking the {\tt DoCon-A 2.00} library needs the minimum of
11 G byte of heap and takes 70 minutes on a 3 GHz personal computer\\
(for Agda 2.5.3, ghc-7.10.2, Debian Linux).

I think that the type check cost currently presents the main problem for {\tt Agda}\\
(it looks like the {\tt Coq} system also has such).

And there are possible and desirable certain optimizations in the type checker, which 
would, probably, reduce the above cost about dozen of times.

\subsubsection{Proof size}

On the example of the {\tt DoCon-A 2.00} library, it occurs that 
the text size of proofs in the source code is approximately five times larger than
the size of the the corresponding textbook containing all the necessary rigorous 
constructive ``humanly'' definitions and proofs.

With further development of the prover tools (in the library part), the proof volume
will become smaller.

\subsection{Examples of what is proved}

Let us list some proof examples implemented in {\tt DoCon-A 2.00}.

\begin{itemize}
\item ``An inverse in a group is unique, and it holds  $(x y)^{ -1} =  y^{ -1} x^{ -1}$''.

\item There are proved various properties of the residue ring \ {\tt Q = R/(b)} \ 
      for any Euclidean ring {\tt R}, depending on the value of {\tt b}.

\item There are proved the base properties of the extended GCD algorithm in any
      Euclidean ring \linebreak ({\tt u a + v b ≈ gcd a b}).

\item It is proved that in any {\tt FactorizationRing} factorization uniqueness is 
      equivalent to the property \ 
      {\footnotesize 
           {\tt Prime∣split =  \ ∀ {p a b} → IsPrime p → p ∣ (a ∙ b) → p ∣ a ⊎ p ∣ b}
      }.
\item It is proved that in any {\tt EuclideanRing} it holds the property {\tt Prime∣split}. 
      The proof uses the above properties of the extended gcd method.
      This brings the unique factorization property to any Euclidean ring with
      factorization, for example, to {\tt Integer}.

\item The correctness of a certain optimized method for summing fractions over a
      domain {\tt R} is derived from the condition of the unique factorization ring for {\tt R}.

\item The last three proofs automatically produce a correctness proof for optimized
      fraction addition over any Euclidean ring with factorization, in particular,
      over {\tt Integer}. 
\end{itemize}

\subsection{Tools to compose proofs}

Usually a programmer ``translates'' a rigorous constructive proof from a textbook or a paper
into a machine certificate.

\emph{It is remarkable that} for the proofs in the current library there are  
sufficient only the three constructs for composing a reasonably looking proof:

\begin{enumerate}
\item normalization,
\item composition of functions,
\item recursion.
\end{enumerate}

\noindent Here
(1) is a proof by computation --- by normalizing to the same term,\\
(2) represents a proof by using a lemma,\\
(3) represents a proof by induction by construction of an argument data.

\subsection{Proof meaning}

Some mathematicians have the following prejudice:\\
``Programs in the verified programming tools (like {\tt Coq, Agda}) do not
provide a proof itself, instead they provide an algorithm to build a proof
witness for each concrete data''.

I claim: \emph{they also provide a proof in its ordinary meaning}\\ 
(this is so in {\tt Agda}, and I expect, the same is with {\tt Coq}).

Let us illustrate this with the example of proving the statement
\begin{center}
                       for all $n \ (n ≤ n)$.
\end{center}

\noindent
for natural numbers. The relation \verb#_≤_# is defined on ℕ so that a witness
for it can be built only with applying the two data constructors (axioms):

{\tt z≤n} \ --- ``{\footnotesize {\tt 0 ≤ n} \ for all {\tt n}}'', \ and \

{\tt s≤s} \ --- ``{\footnotesize if {\tt m ≤ n}, \ then \ {\tt suc m ≤ suc n}}''.
\\
(Syntax: \ {\footnotesize {\tt z≤n, s≤s}} \ are function names, as they are written 
                                                                   without blanks).

For example: \ {\footnotesize {\tt s≤s (s≤s z≤n)}} \ is a proof for the statement \
{\footnotesize {\tt 2 ≤ 5}}.

Consider the inductive proof for the goal statement.

If {\footnotesize {\tt n = 0}}, then {\footnotesize {\tt 0 ≤ 0}} is proved by the law 
{\tt z≤n}.
For a nonzero, it is needed to prove \ {\footnotesize {\tt suc n ≤ suc n}}. \
By the inductive supposition, there exists a proof  {\tt p}  for
{\footnotesize {\tt n ≤ n}}. And the law  {\tt s≤s}  applied to  {\tt p}
yields a proof for \ {\footnotesize {\tt suc n ≤ suc n}}.

The corresponding proof in {\tt Agda} is
{\footnotesize
\begin{code}
  theorem :  ∀ n → n ≤ n
  theorem 0       =  z≤n
  theorem (suc n) =  s≤s (theorem n)
\end{code}
}
\noindent
For each {\footnotesize {\tt n : ℕ}} \ the function
{\footnotesize {\tt theorem}} returns a value in the type
{\footnotesize {\tt n ≤ n}}, \ that is the corresponding witness.

The second pattern applies the function {\footnotesize {\tt theorem}}
recursively.
This all provides a \emph{proof in the two meanings}.
\
(1) At the run-time, {\footnotesize {\tt (theorem n)}} yields a proof for \
    {\footnotesize {\tt n ≤ n}} \ for each concrete {\tt n}. \
(2) The very algorithm expression for {\footnotesize {\tt theorem}} is a
    symbolic expression that presents a general proof for the statement
    {\footnotesize ``for all {\tt n (n ≤ n)}''}.

The algorithm (program) {\footnotesize {\tt theorem}} is a symbolic expression
(term), its parts depending on a variable {\tt n}. This term is verified by
the type checker statically --- before run-time.
And this is the same as checking an ordinary inductive proof.
Reasoning by induction corresponds to setting a recursive call to the algorithm for 
forming a witness.

We see that (2) provides a real generic proof for the statement, while (1)
provides a witness for each concrete {\tt n}.
Similar it is with all proofs.

\subsection{The problem of setting proofs}

\subsubsection{About proof translation}

Many proofs in the current library have been obtained from known rigorous 
constructive humanly proofs by ``translation'' to {\tt Agda}. And the time and 
effort for this translation occur somewhat 50 times greater than I expected 
(this depends on the skill of a person, though).
This contradicts to our expectation of that the above translation needs to be more 
or less a mechanical procedure.
This presents a certain question for us.

\subsubsection{Not only translation}

In the literature we often meet non-constructive proofs that can be replaced with 
constructive ones. This replacement often needs a nontrivial invention. And such an
invention is currently done by an human much simpler than by any prover, a prover
will not help, at the current state of art.

Also even ``rigorous'' constructive proofs in literature usually have considerable 
gaps; these gaps can be filled by the reader in a way more or less evident to the reader 
and to the author. For a machine-checked proof, someone needs to fill these gaps
with machine-checked proofs, and most often this filling is not automatic.

\subsubsection{Libraries of lemmata}

Currently the main tool that helps composing proofs is 
\emph{accumulating the library of lemmata}. This follows an usual approach to developing
a theory in mathematics.

For example, the library {\tt AList} provides proofs for many general-purpose 
lemmata for association lists. In particular, they help to prove that the multiset sum
satisfies the laws of associativity and commutativity, and this is used further in
operations with the factorization data structures.

\subsubsection{Special provers}

At the current state of art, an automatic prover can be really useful in special problems, 
where it is known an algorithm for solving a problem. This frees the programmer from 
setting manually a great number of proof steps.
For example, many {\tt Agda} proofs for equalities in the {\tt DoCon-A} library can be 
automatically reduced to applying the Gr\"obner basis algorithm 
(the method from \cite{CA-p} Appendix I, modified for integer coefficients as shown in
\cite{Me3}). 
Though, it needs to be designed a certain translator, similar to the translators in 
{\tt EqProver/*}, but a more complex one.

For a more generic case, there can be applied various versions of the 
\emph{completion method} (a part of the \emph{term rewriting} theory) for deriving 
an equality from other equalities --- even though it is a semidecision procedure.

\subsubsection{About general provers}

But special provers cover (on average) may be only 1/3 part of the proof design effort.
I do not expect that applying the existing \emph{generic} provers can make it any 
essentially easier. This is for the following reason.
Speaking informally, searching for a proof in an {\tt Agda} program consists of the 
following parts.
\begin{enumerate}
\item Inventing replacement for non-constructive parts.

\item Breaking the goal to several lemmata.

\item Applying a fixed finite set of standard proof attempts 
      (induction by the chosen value, considering cases for the chosen value, and some 
       others), thus developing a search tree of attempts. 

\item Applying special provers to the appropriate parts.
\end{enumerate}

And it occurs so that a prover often helps essentially only in the part (4).
 
The parts (1) and (2) are done much simpler by an human.

Consider the part (3). Choosing the current attempt to try is done much better 
by an human. This also concerns choosing the right expression to apply induction on,
choosing the right expression to apply considering cases for, and so on ---
all this is done much easier by an human.

It remains the part (4). Here special provers help a lot.

So that the situation is: the automatic part (4) covers about 1/3 of the whole formal 
proof invention effort (assuming that other parts often include sub-parts done by (4)),
and for all the rest, provers do not help any essentially.

\subsubsection{Example}

Consider the function {\tt rev} for reversing a list, where {\tt rev} is implemented via 
repeated concatenation {\tt ++}. The goal is to prove that 
{\tt (rev ∘ rev)}  is the identic map. Represent this theorem by the function {\tt revrev}.
 
First the programmer needs to search for an humanly proof.
It is natural to try a proof by induction by the construction of the argument list {\tt xs}.
The step of induction has the goal of deriving the equality \ \ 
{\footnotesize {\tt rev (rev (x ∷ xs)) ≡ x ∷ xs}} \ \ from the equality \ \  
{\footnotesize {\tt (1): \ rev xs ≡ xs}}.\linebreak
The goal equality is normalized to the goal \ \
{\footnotesize {\tt rev ((rev xs) ++ [x]) ≡ x ∷ xs}}.

This is a difficult point for a prover (an automatic one, or an human). Automatic provers
usually continue to develop the search tree by various attempts. There is a small 
finite set of derivation rule kinds, one of such rules is induction by some chosen value. 
This leads a prover to infinity --- unless it ``guesses'' to apply at this point 
\emph{searching a lemma}. A lemma needs to be some statement {\tt L} such that {\tt L}
is proved by the prover in a reasonable number of steps, and then, the goal is derived 
successfully from {\tt L} (this needs several recursive calls of the prover, with giving
it a bound for the number of the search steps).

The human intuition hints that this lemma needs to be some equality for the term \linebreak 
{\footnotesize {\tt rev ((rev xs) ++ [ x ])}} \ \ in the goal. The intuition also hints to 
choose the equality \linebreak {\footnotesize {\tt rev (ys ++ [ x ]) ≡ x ∷ (rev ys)}}, \ \ 
and further, to substitute {\footnotesize {\tt (rev xs)}} for the variable {\tt ys}. 
Let us call this lemma {\tt rev-append}. This lemma is easy to prove by induction by 
{\tt ys}.
And this lemma fills the gap in the goal proof, it only remains to apply the goal 
statement recursively:\linebreak {\footnotesize {rev (rev xs) ≡ xs}}.

Translating the above found humanly proof to {\tt Agda} yields the following program\\
(this is also an example of setting proofs in an {\tt Agda} program):

{\footnotesize
\begin{code}
-------------------------------------------------------------------------------------
open import Relation.Binary.PropositionalEquality as PE using (_≡_)
open import Data.List using (List; []; _∷_; [_])
open PE.≡-Reasoning renaming (_≡⟨_⟩_ to _≡[_]_; begin_ to ≡begin_;  _□ to _≡end)

module _ {a} (A : Set a) 
  where
  _++_ : List A → List A → List A     -- concatenation
  []       ++ ys =  ys
  (x ∷ xs) ++ ys =  x ∷ (xs ++ ys)

  rev : List A → List A               -- reversing a list
  rev []       = []
  rev (x ∷ xs) = (rev xs) ++ [ x ] 

  rev-append :  ∀ x ys →  rev (ys ++ [ x ]) ≡ x ∷ (rev ys)       -- lemma

  rev-append x []       =  PE.refl
  rev-append x (y ∷ ys) =  
      ≡begin
        rev ((y ∷ ys) ++ [ x ])        ≡[ PE.refl ]       
        rev (y ∷ (ys ++ [ x ]))        ≡[ PE.refl ]
        (rev (ys ++ [ x ])) ++ [ y ]   ≡[ PE.cong (_++ [ y ]) (rev-append x ys) ]
        (x ∷ (rev ys)) ++ [ y ]        ≡[ PE.refl ]
        x ∷ ((rev ys) ++ [ y ])        ≡[ PE.refl ]
        x ∷ (rev (y ∷ ys))
      ≡end

  revrev :  ∀ xs → rev (rev xs) ≡ xs                   -- the goal theorem
  revrev []       =  PE.refl
  revrev (x ∷ xs) =  ≡begin rev (rev (x ∷ xs))         ≡[ PE.refl ]
                            rev ((rev xs) ++ [ x ])    ≡[ rev-append x (rev xs) ]
                            x ∷ (rev (rev xs))         ≡[ PE.cong (x ∷_) (revrev xs) ]
                            x ∷ xs
                     ≡end
--------------------------------------------------------------------------------------
\end{code}
}

\noindent 
Here the {\tt PE.refl} proof step means (according to the general step of the proof 
by normalization) applying normalization of a term by the equations of the function 
{\tt rev}.

How could an automatic prover help essentially in composing the above prove?

All the points are easy for the programmer, except searching for a lemma.
But this part is done much easier by an human.

There are some provers based on many-sorted term rewriting which include the step of searching
a lemma in the form of equality. There is a method for rejecting fast most of useless lemma 
candidates. Still the cost of the search--through is enormous. 
One of such provers finds an useful lemma after 10 minutes (on a 3 GHz machine). 
Still this part is easier for an human. 
Also this success depends on the initial choice of the operator set detected as related to 
the goal (this is a certain heuristic).
This luck is not stable, and in most real examples an automatic prover will not help.

Summing it up: provers are useful in the part (4) of special provers --- which is essential, 
but a large part of (1), (2), (3) is practically unaccessible for automatic provers
--- at the current state of art.

The problem of automating these parts (as possible) is principal one for the area of provable 
programming, and the most difficult one.

\end{document}